\documentclass[twocolumn,showpacs,preprintnumbers,amsmath,amssymb,superscriptaddress]{revtex4}

\usepackage{graphicx}
\usepackage{dcolumn}
\usepackage{bm}

\begin{document}
\draft
\title{True ternary fission of superheavy nuclei}

\author{V.I.~Zagrebaev}
\affiliation{Flerov Laboratory of Nuclear Reactions, JINR, Dubna,
Moscow Region, Russia}
\author{A.V.~Karpov}
\affiliation{Flerov Laboratory of Nuclear Reactions, JINR, Dubna,
Moscow Region, Russia}
\author{Walter~Greiner}
\affiliation{Frankfurt Institute for Advanced Studies, J.W.
Goethe-Universit\"{a}t, Frankfurt, Germany}

\date{\today}

\begin{abstract}
We found that a true ternary fission with formation of a heavy third
fragment (a new type of radioactivity) is quite possible for
superheavy nuclei due to the strong shell effects leading to a
three-body clusterization with the two doubly magic tin-like
cores. The simplest way to discover this phenomenon in the decay of
excited superheavy nuclei is a detection of two tin-like
clusters with appropriate kinematics in low-energy collisions of
medium mass nuclei with actinide targets. The three-body
quasi-fission process could be even more pronounced for giant
nuclear systems formed in collisions of heavy actinide nuclei. In
this case a three-body clusterization might be proved
experimentally by detection of two coincident lead-like fragments
in low-energy U+U collisions.
\end{abstract}
 \maketitle

\begin{figure}[ht]
\begin{center} 
\includegraphics[width = 6.0 cm, bb=0 0 969 804]{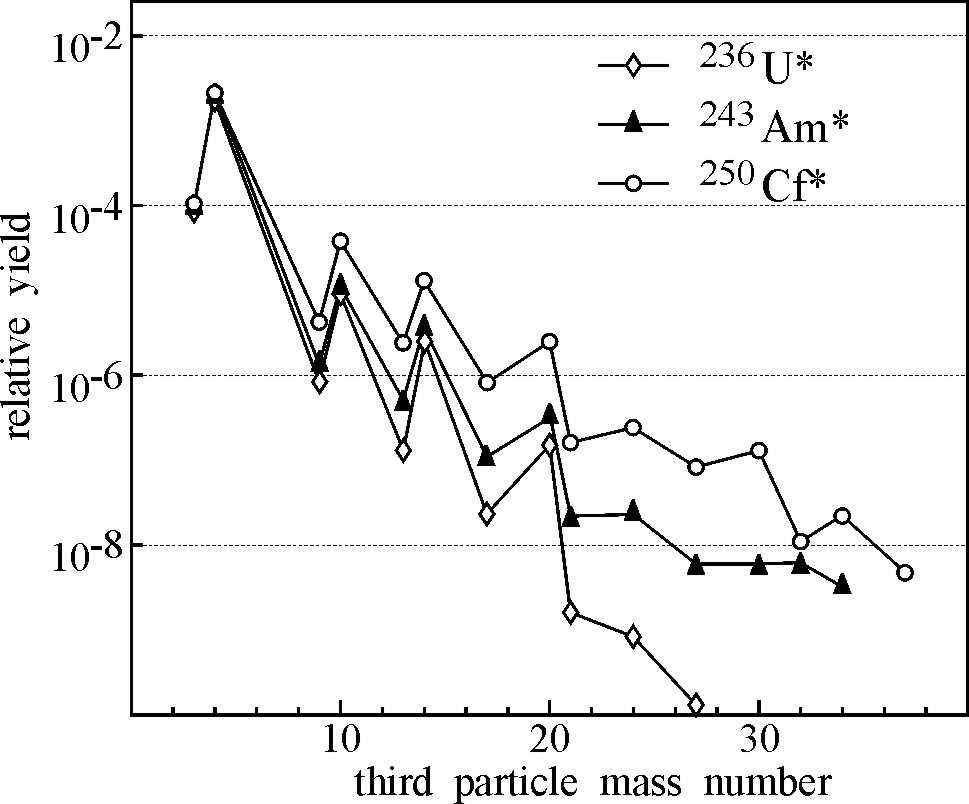} \end{center}
\caption{Relative to binary fission yields of ternary particles in the $(n_{\rm th},f)$ reactions with
thermal neutrons \cite{Gonnenwein99}.\label{lp_ternary}}
\end{figure}

Today the term ``ternary fission'' is commonly used to denote the
process of formation of light charged particle accompanied fission
\cite{Vag91}. This is a rare process (less than 1\%) relative to
binary fission, see Fig.\ \ref{lp_ternary}. As can be seen the
probability of such a process decreases sharply with increasing mass
number of the accompanied third particle. These light particles are
emitted almost perpendicularly with respect to the fission axis
(equatorial emission)  \cite{Vag91}. It is interpreted as an
indication that the light ternary particles are emitted from the
neck region and are accelerated by the Coulomb fields of both
heavy fragments.

In contrast to such a process, the term ``true ternary fission''
is used for a simultaneous decay of a heavy nucleus into three
fragments of not very different mass \cite{Vag91}. Such decays of
low excited heavy nuclei were not observed yet. The true ternary
fission of atomic nuclei (below we omit the word ``true'') has a
long history of theoretical and experimental studies. Early
theoretical considerations based on the liquid drop model (LDM)
\cite{Sw58} showed that for heavy nuclei ternary fission produces
a larger total energy release in comparison to binary fission, but
the actual possibility of ternary fission is decided, in fact, by
barrier properties and not by the total energy release. It was
found that the LDM ternary fission barriers for oblate (triangle)
deformations are much higher as compared to the barriers of
prolate configurations \cite{Diehl74}, and it seems that the
oblate ternary fission may be excluded from consideration. However
further study of this problem within the more sophisticated
three-center shell model \cite{Maruhn79} showed that the shell
effects may significantly reduce the ternary fission barriers even
for oblate deformations of very heavy nuclei.

It is well known that for superheavy nuclei the LDM fission
barriers are rather low (or vanish completely) and the shell
correction to the total deformation energy is very important.
First estimations of the binary and prolate ternary fission
barriers of superheavy nucleus $^{298}$114, made in
\cite{Schult74} with the shell corrections calculated in an
approximate way, demonstrated that they are identical to within 10\%. To our
knowledge, since then there was not any significant progress in
theoretical (or experimental) study of ternary fission. In the
meanwhile, today it becomes possible to study experimentally the
properties and dynamics of formation and decay of superheavy
nuclei \cite{Ogan07}, for which the ternary fission could be
rather probable (see below).

The two-center shell model (TCSM) \cite{TCSM} looks most
appropriate for calculation of the adiabatic potential energy of
heavy nucleus at large dynamic deformations up to the configuration of
two separated fragments. The nuclear shape in this model is
determined by 5 parameters: the elongation $R$ of the system,
which for separated nuclei is the distance between their mass
centers; the ellipsoidal deformations of the two parts of the
system $\delta_1$ and $\delta_2$; the mass-asymmetry parameter
$\eta = (A_2-A_1)/(A_2 + A_1)$, where $A_1$ and $A_2$ are the mass
numbers of the system halves; and the neck parameter $\epsilon$
which smoothes the shape of overlapping nuclei.

Within the macro-microscopic approaches the energy of the deformed nucleus is composed of the two parts
$E(A,Z; R, \delta, \eta, \epsilon) = E_{\rm mac}(A,Z; R, \delta, \eta,\epsilon) + \delta E(A,Z; R, \delta, \eta,\epsilon)$.
The macroscopic part, $E_{\rm mac}$, smoothly depends on the proton and neutron numbers and may be calculated within the LDM.
The microscopic part, $\delta E$, describes the shell effects.
It is constructed from the single-particle energy spectra by the Strutinsky procedure \cite{Strutinsky}.
The details of calculation of the single particle energy spectra within the TCSM, the explanation of all the parameters used
as well as the extended and empirical versions of the TCSM  may be found in \cite{ZKAG07}.

\begin{figure}[h]\begin{center}
\includegraphics[width = 8.5 cm, bb=0 0 1294 248]{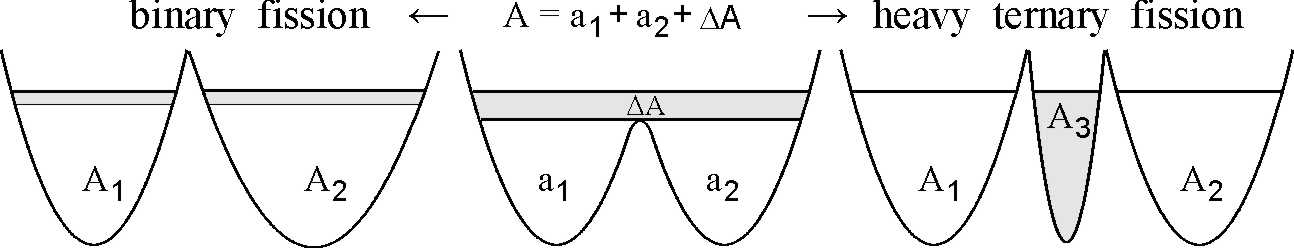}\end{center}
\caption{Schematic view of binary and ternary
fission.\label{ternary}}
\end{figure}

Within the TCSM for a given nuclear configuration ($R, \eta,
\delta_1,\delta_2$) we may unambiguously determine the two
deformed cores $a_1$ and $a_2$ surrounded with a certain number of
shared nucleons $\Delta A=A_{\rm CN}-a_1-a_2$ (see Fig.\
\ref{ternary}). During binary fission these valence nucleons
gradually spread between the two cores with formation of two final
fragments $A_1$ and $A_2$. Thus, the processes of compound nucleus
(CN) formation, binary fission and quasi-fission may be described
both in the space of the shape parameters ($R, \eta,
\delta_1,\delta_2$) and in the space
($a_1,\delta_1,a_2,\delta_2$). This double choice of equivalent
sets of coordinates is extremely important for a clear understanding
and interpretation of the physical meaning of the intermediate local
minima appearing on the multi-dimensional adiabatic potential
energy surface and could be used for extension of the model for
description of three-core configurations appearing in ternary
fission.

\begin{figure}[h]
\begin{center} \includegraphics[width = 6.5 cm, bb=0 0 954 2250]{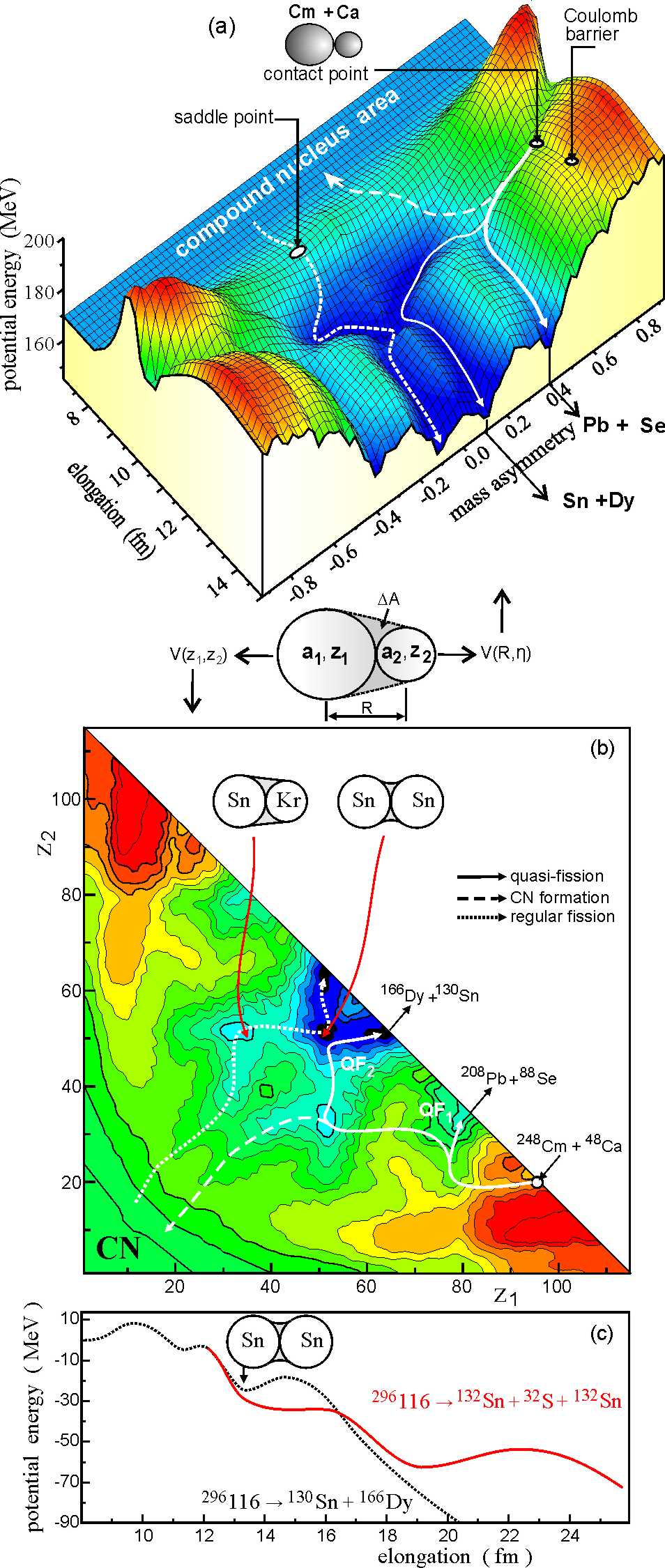} \end{center}
\caption{Adiabatic potential energy for nucleus $^{296}$116 formed in collision of $^{48}$Ca with
$^{248}$Cm. (a) Potential energy in the ``elongation-–mass asymmetry'' space. (b) Topographical landscape
of the same potential in the ($z_1, z_2$) plane. Dashed, solid and dotted curves show most
probable trajectories of fusion, quasi-fission and regular fission, respectively. The diagonal corresponds
to the contact configurations ($R = R_{\rm cont}, z_1+z_2 = Z_{\rm CN}, \Delta A = 0$).
(c) Potential energy calculated for binary (dotted curve) and symmetric ternary fission
of nucleus $^{296}$116.\label{vfusfis}}
\end{figure}

The adiabatic driving potential for formation and decay of the superheavy nucleus $^{296}$116 at fixed deformations of both
fragments is shown in Fig.\ \ref{vfusfis} as a function of elongation and mass asymmetry and also as a function of charge
numbers $z_1$ and $z_2$ of the two cores (minimized over neutron numbers $n_1$ and $n_2$) at $R\le R_{\rm cont}$.
Following the fission path (dotted curves in Fig.\ \ref{vfusfis}a,b) the nuclear system passes through the
optimal configurations (with minimal potential energy) and overcomes the multi-humped fission barrier.
The intermediate minima located along this path correspond to the shape isomeric states.
These isomeric states are nothing else but the two-cluster configurations with magic or semi-magic cores
surrounded with a certain amount of shared nucleons.
In the case of binary fission of nucleus $^{296}$116 the second (after ground state) minimum on the fission path
arises from the two-cluster nuclear configuration consisting of tin-like ( $z_1=50$) and krypton-like ( $z_2=36$) cores
and about 70 shared nucleons. The third minimum corresponds to the mass-symmetric clusterization with two magic
tin cores surrounded with about 30 common nucleons.

A three-body clusterization might appear just on the path from the
saddle point to scission, where the shared nucleons $\Delta A$ may
form a third fragment located between the two heavy clusters $a_1$
and $a_2$. In Fig.~\ref{ternary} a schematic view is shown for
binary and ternary fission starting from the configuration of the
last shape isomeric minimum of CN consisting of two magic tin
cores and about 30 extra (valence) nucleons shared between the two
clusters and moving initially in the whole volume of the
mono-nucleus. In the case of two-body fission of $^{296}$116
nucleus these extra nucleons gradually pass into one of the
fragments with formation of two nuclei in the exit channel (Sn and
Dy in our case, see the fission path in Fig.~\ref{vfusfis},
mass-symmetric fission of $^{296}$116 nucleus is less favorable).
However there is a chance for these extra nucleons $\Delta A$ to
concentrate in the neck region between the two cores and form
finally the third fission fragment.

There are too many collective degrees of freedom needed for proper
description of the potential energy of a nuclear configuration
consisting of three deformed heavy fragments. We restricted
ourselves by consideration of the potential energy of a three-body
symmetric configuration with two equal cores $a_1=a_2$ (and, thus,
with two equal fragments $A_1=A_2$ in the exit fission channels).
Also we assume equal dynamic deformations of all the fragments,
$\delta_1=\delta_2=\delta_3=\delta$, and use the same shape
parametrization for axially symmetric ternary fission as in
Ref.~\cite{Wu84} (determined by three smoothed oscillator
potentials).

The third fragment, $a_3$, appears between the two cores when the
total elongation of the system, described by the variable $R$
(distance between $a_1$ and $a_2$), is sufficiently large to
contain all three fragments, i.e., $R\ge R(a_1)+2R(a_3)+R(a_2)$.
Finally, we calculated the three-dimensional potential energy
$V(R,\delta,A_3)$ trying to find a preferable path for ternary
fission and estimate how much larger the barrier is
for three-body decay as compared to binary fission. For better visualization we
plot the calculated potential energy $V(R,\delta,A_3)$ as a
function of $(R/R_0-1)\cos{(\alpha_3)}$ and
$(R/R_0-1)\sin{(\alpha_3)}$ at fixed dynamic deformation
$\delta=0.2$, where $\alpha_3 = \pi\cdot A_3/100$ and $R_0$ is the
radius of sphere of equivalent volume (CN).

\begin{figure}[h]
\begin{center}\includegraphics[width = 6.7 cm, bb=0 0 882 929]{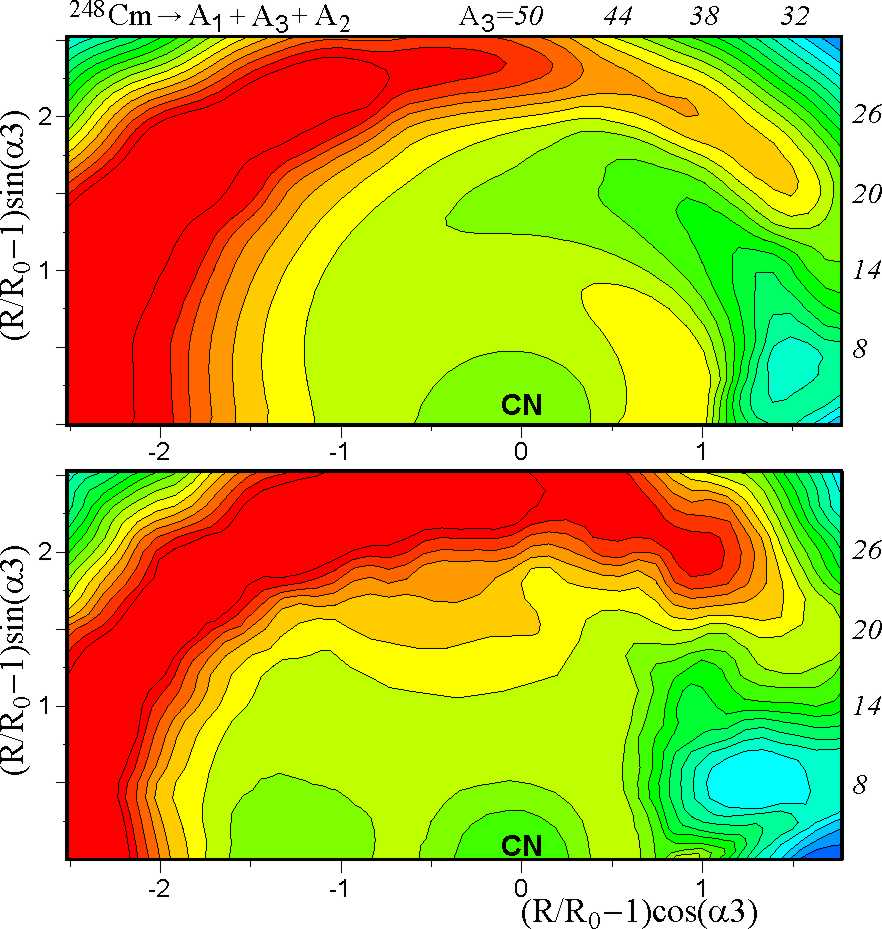} \end{center}
\caption{Potential energy for ternary fission of $^{248}$Cm.
Macroscopic part of potential energy and the total one (LDM plus
shell corrections) are shown at upper and bottom panels,
respectively, depending on elongation and mass of third fragment
(italic numbers). Contour lines are drawn over 3~MeV.
\label{Cm248}}
\end{figure}

The macroscopic (LDM) part of the potential energy for $^{248}$Cm
is shown in upper panel of Fig.~\ref{Cm248}. The binary fission of
$^{248}$Cm evidently dominates because after the barrier the
potential energy is much steeper just in the binary exit channel
(right bottom corner, $A_3\sim 0$). Emission of light third
particle is possible here but not the true ternary fission. The
shell correction (which makes deeper the ground state of this
nucleus by about 3~MeV) does not change distinctively the total
potential energy (see the bottom panel of Fig.~\ref{Cm248}).
Nevertheless the experiments aimed on the observation of real ternary
fission of actinide nuclei (with formation of heavy third
fragment) are currently in progress \cite{Kamanin08}.

\begin{figure}[h]
\begin{center} \includegraphics[width = 6.7 cm, bb=0 0 882 932]{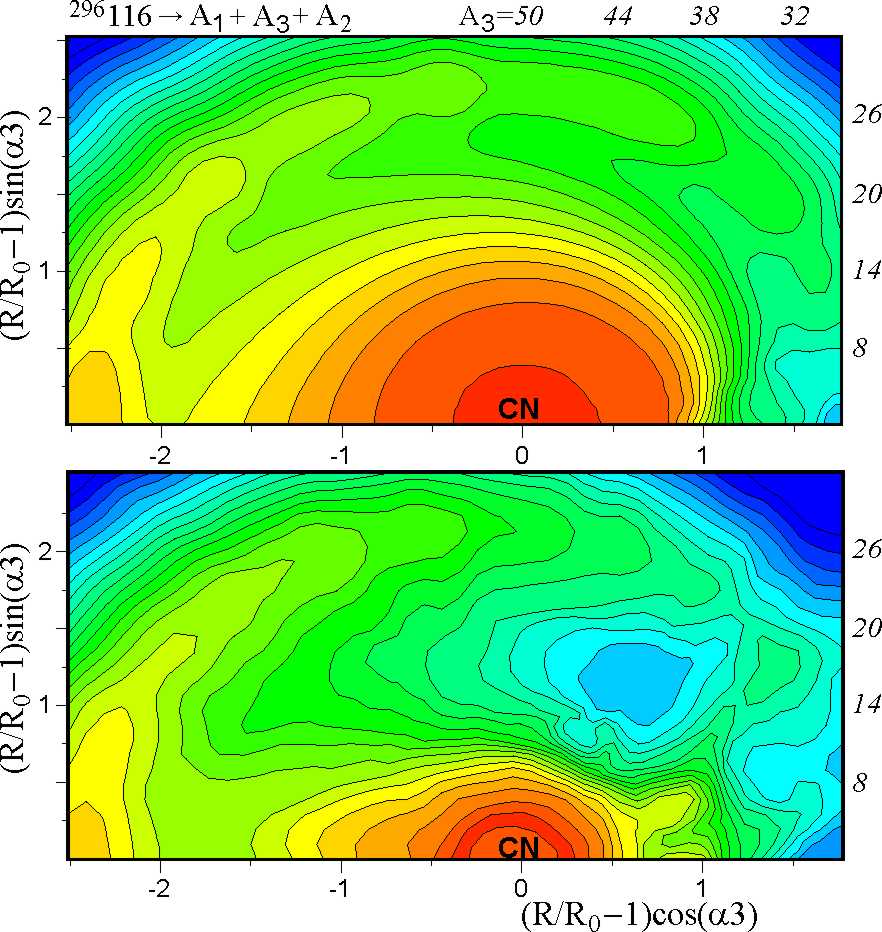} \end{center}
\caption{The same as Fig.~\ref{Cm248} but for superheavy nucleus
$^{296}$116. Contour lines are drawn over 5~MeV.\label{116}}
\end{figure}

In the case of superheavy nuclei the macroscopic potential energy
does not lead to any barrier at all (neither in binary nor in
ternary exit channel) and stability of these nuclei is determined
completely by the shell corrections. In Fig.~\ref{116} the
calculated potential energy is shown for superheavy nucleus
$^{296}$116. In contrast with $^{248}$Cm, in this case a real
possibility for ternary fission appears with formation of third
fragment $A_3\sim 30$ and two heavy fragments $A_1=A_2 \sim 130$.
The ternary fission valley is quite well separated by the
potential ridge from the binary fission valley. This means that
the ternary fission of $^{296}$116 nucleus into the
``tin--sulfur--tin'' combination should dominate as compared with
other true ternary fission channels of this nucleus.

More sophisticated consideration of the multi-dimensional
potential energy surface is needed to estimate the ``ternary
fission barrier'' accurately. However, as can be seen from
Fig.~\ref{116}, the height of the ternary fission barrier is not
immensely high. It is quite comparable with the regular fission
barrier because the ternary fission starts in fact from the
configuration of the shape isomeric state which is located outside
from the first (highest) saddle point of superheavy nucleus
$^{296}$116 (see the solid curve on the bottom panel of
Fig.~\ref{vfusfis}).

\begin{figure}[h]\begin{center} \includegraphics[width = 8.4 cm, bb=0 0 2416 1089]{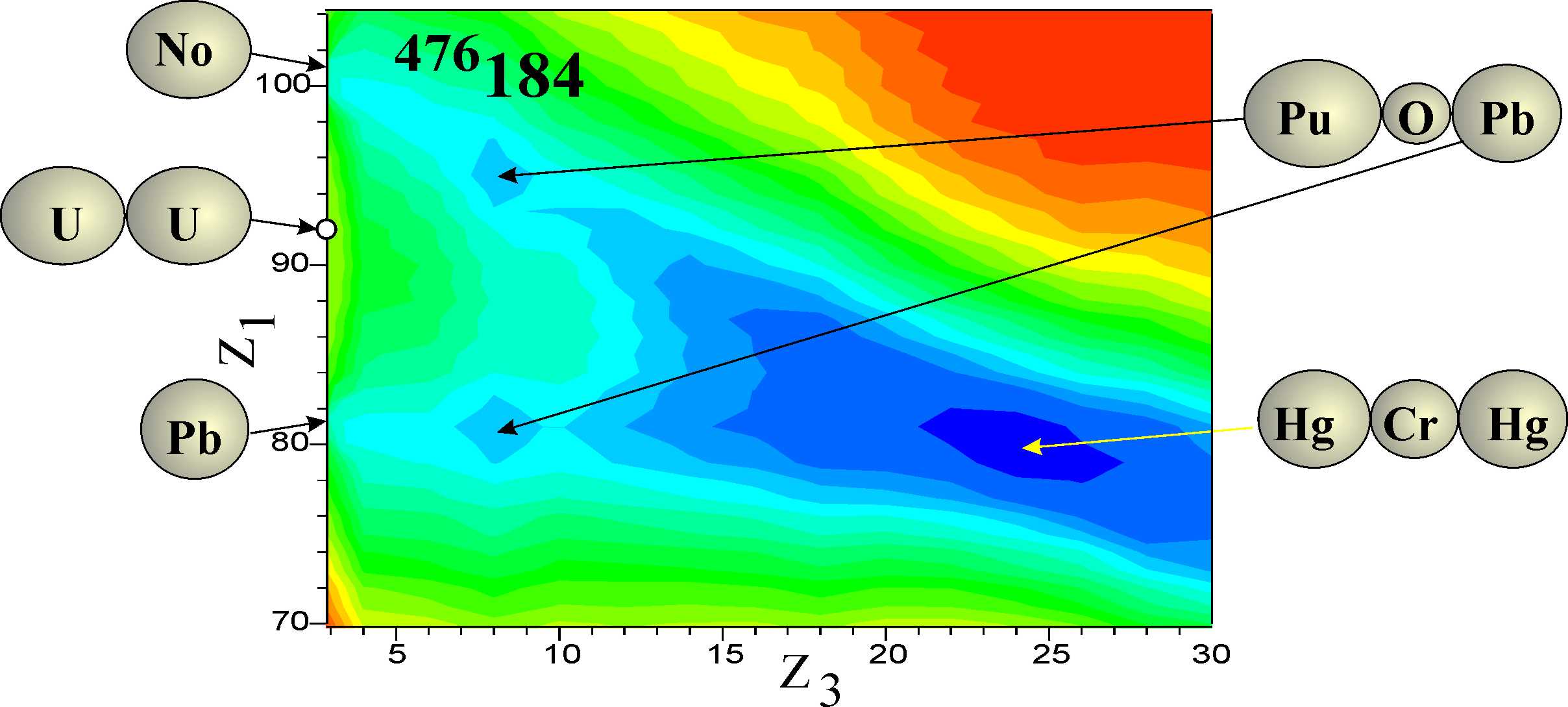} \end{center}
\caption{Landscape of potential energy of three-body
configurations formed in collision of
$^{238}$U+$^{238}$U.\label{3clustersUU}}
\end{figure}

\begin{figure}[h]\begin{center} \includegraphics[width = 6.0 cm, bb=0 0 1824 1179]{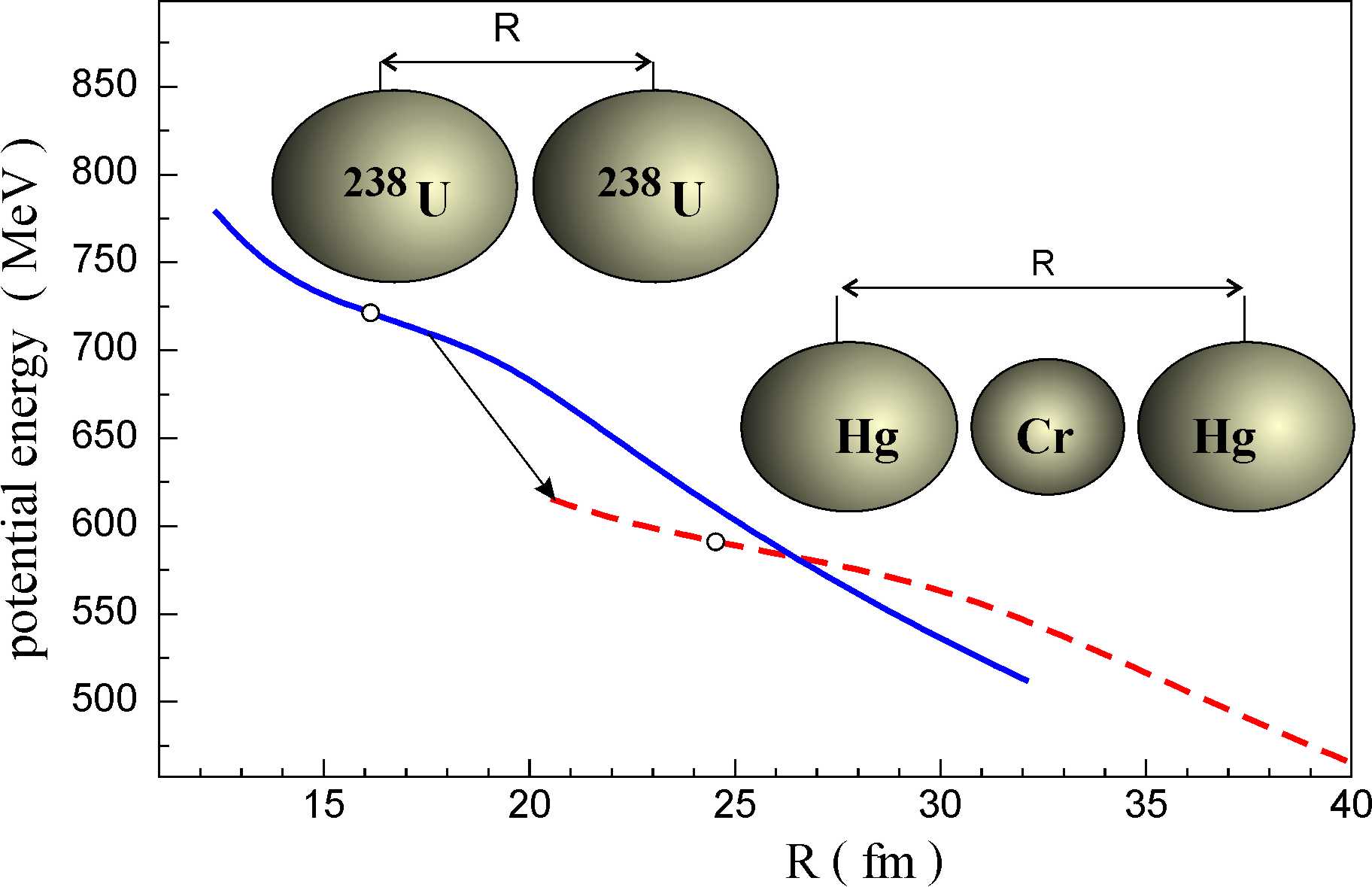} \end{center}
\caption{Radial dependence of the potential energy of two uranium
nuclei (solid curve) and of the three-body nuclear configuration
formed in collision of $^{238}$U+$^{238}$U (dashed
curve).\label{3clustersR}}
\end{figure}

Conditions for the three-body fission (quasi-fission) are even
better in the giant nuclear systems formed in low-energy
collisions of actinide nuclei. In this case the shell effects
significantly reduce the potential energy of the three-cluster
configurations with two strongly bound lead-like fragments. In
Fig.~\ref{3clustersUU} the landscape of the potential energy
surface is shown for a three-body clusterization of the nuclear
system formed in collision of U+U. Here the potential energy was
calculated as a function of three variables, $Z_1$, $Z_3$ and $R$
at fixed (equal) deformations of the fragments being in contact
($R_1+2R_3+R_2=R$). To make the result quite visible we minimized
the potential energy over the neutron numbers of the fragments,
$N_1$ and $N_3$.

As can be seen, the giant nuclear system, consisting of two
touching uranium nuclei, may split into the two-body exit channel
with formation of lead-like fragment and complementary superheavy
nucleus (the so-called anti-symmetrizing quasi-fission process
which may lead to an enhanced yield of SH nuclei in multi-nucleon
transfer reactions \cite{Zag06}). Beside the two-body Pb--No
clasterization and the shallow local three-body minimum with
formation of light intermediate oxygen-like cluster, the potential
energy has the very deep minimum corresponding to the
Pb-Ca-Pb--like configuration (or Hg-Cr-Hg) caused by the N=126 and
Z=82 nuclear shells.

Thus we found that for superheavy nuclei the three-body
clusterization (and, hence, real ternary fission with a heavy
third fragment) is quite possible. The simplest way to discover
this phenomenon is a detection of two tin or xenon-like clusters in
low energy collisions of medium mass nuclei with actinide targets,
for example, in $^{64}$Ni+$^{238}$U reaction. These unusual decays
could be searched for also among the spontaneous fission events of
superheavy nuclei \cite{Ogan07}.

The extreme clustering process of formation of two lead-like
double magic fragments in collisions of actinide nuclei is also a
very interesting subject for experimental study. Such
measurements, in our opinion, are not too difficult. It is
sufficient to detect two coincident lead-like ejectiles (or one
lead-like and one calcium-like fragments) in U+U collisions to
conclude unambiguously about the ternary fission of the giant
nuclear system. More flat radial dependence of the potential
energy (as compared with a two-body system) is another feature of
the three body clusterization, see Fig.~\ref{3clustersR}. This
means that decay of U+U--like nuclear system into the
energetically preferable (and more stable in some sense)
three-body configurations may also significantly prolong the reaction
time, which (among other things) could be important for
spontaneous positron formation in super-strong electric field.

We are indebted to the DFG -- RFBR collaboration for support of
our studies.

\end{document}